\documentclass[twocolumn,showpacs,preprintnumbers,amsmath,amssymb]{revtex4}

\usepackage{amsmath}
\usepackage{amssymb}
\usepackage{amsfonts}
\usepackage{graphicx}
\usepackage{dcolumn}
\usepackage{bm}
\usepackage{epsfig}

\begin{document}

\newcommand{\bc}{\begin{center}}
\newcommand{\ec}{\end{center}}
\newcommand{\be}{\begin{equation}}
\newcommand{\ee}{\end{equation}}
\newcommand{\beqn}{\begin{eqnarray}}
\newcommand{\eeqn}{\end{eqnarray}}

\title{Impact of short- and long-range forces on protein conformation and adsorption kinetics}

\author{Anthony Quinn}
\altaffiliation[Present address: ]{Department of Chemical \& Biomolecular Engineering, University of Melbourne, Australia}
\author{Hubert Mantz}
\author{Karin Jacobs}
\affiliation{
  Department of Experimental Physics, Saarland University, 66041 Saarbr\"ucken, Germany
}
\author{Markus Bellion}
\author{Ludger Santen}
\affiliation{
Department of Theoretical Physics, Saarland University, 66041 Saarbr\"ucken, Germany
}

\begin{abstract}
We have studied the adsorption kinetics of the protein amylase at
solid/liquid interfaces. Offering substrates with tailored properties, we are
able to separate the impact of short- and long-range
interactions. By means of a colloidal Monte Carlo approach including 
conformational changes of the adsorbed proteins induced by density 
fluctuations, we develop a scenario that is consistent with the
experimentally observed three-step kinetics on specific substrates.
Our observations show that not only the surface chemistry determines the 
properties of an adsorbed protein layer but also the van der Waals 
contributions of a composite substrate may lead to non-negligible effects.
\end{abstract}


\maketitle

\section*{Introduction}

The adsorption of proteins at interfaces plays a crucial role in
determining the function of many biological systems, and hence it is
the focus of research activities in chemical and medical
applications. Adsorption is a thermodynamic process that occurs 
spontaneously whenever
protein-containing aqueous solutions contact solid surfaces, and results in 
a modification of the sorbent surface and often that of the protein as 
well~\cite{engel2004}.
The structure of a protein is closely connected to its function and efficacy, 
therefore protein adsorption and surface-induced
conformational changes are important issues in biocompatibility of materials
and have been subject to numerous studies 
(cf.\ \cite{malmsten2003,gray2004,garcia, garcia2} and references therein).
The adsorption of proteins from aqueous solutions is driven by
short- and long-range interactions. The main players in the
latter are van der Waals and Coulomb contributions. Due to the strong
screening of the Coulomb interactions,
the question arises what actually is the leading contribution of these two.
It is the aim of this paper to study the role of the long-range potential 
forces and possible conformational changes for the complex process of
protein adsorption~\cite{norde1986,haynes1994}. 

Studies of {\it in-situ} biofilm formation are an experimental and theoretical 
challenge. The experimental technique must provide a sub-nm spatial 
resolution in normal sample direction as well as a time resolution in 
the range of seconds. Additionally, the sample is not  
accessible directly, rather it is immersed in a liquid and only minute amounts of 
material can be analyzed. These constraints rule out many common thin 
film characterization techniques. Surface plasmon resonance spectroscopy 
(SPR) or the frequency shift of a quartz crystal microbalance upon 
material adsorption will provide a high sensitivity, 
yet both methods suffer from a constraint 
concerning the substrate material. Therefore, ellipsometry seemed 
to us to be the method of choice for the {\it in-situ} studies.
From the viewpoint of modelling, the atomistic simulation of a complex 
biofilm with at least hundreds of mutual interacting macromolecules, each 
of which with its own complexity, is not possible with state-of-the-art 
computers.
Even taking into account the fast development of computational power, 
the time scale of the adsorption process, which is of the order of minutes,
will not be accessible in the next few decades. 
Therefore, the only route to achieve a comprehensive description is
to use a largely simplified 
protein model that has a level of sophistication in accordance with 
the experimental characteristics.

In this study, we combine an experimental and a modelling approach, each 
of which is not able to offer a complete picture of the 
protein adsorption kinetics. In the ellipsometry studies a systematic 
variation of the potential forces
is provided by using tailored composite substrates.
The theoretical investigations are performed by
means of Monte Carlo (MC) simulations utilizing an effective particle
model. The special focus of the MC investigation is on the relation
between conformational changes of the (model) proteins and the adsorption 
kinetics.
By the combination of both methods, we have been able to devise  
a consistent scenario for the biofilm adsorption. Our findings underline the 
importance of long-range forces as well as of induced conformational changes 
of the proteins.

\section*{Experiments}

The protein under investigation, $\alpha$-amylase from human saliva, 
has particular relevance in dental research due to its function in the primary
colonization of bacteria leading to plaque formation~\cite{rogers2001}.
As substrate a composite material was used. Silicon wafers
with natural (2~nm) and thick (192~nm) silicon oxide surfaces cause identical 
short-range interactions with the adsorbate, since the chemical composition 
of the surface is identical~\cite{karin_ralf2001,gecko2005}. However, comparing 
the long-range van der Waals forces acting between adsorbate and oxide 
layer with those acting between adsorbate and bulk material through the oxide
layer, the strength and sign of the Hamaker constant can be different~\cite{karin_ralf2001}.
Varying the oxide layer thickness enables the strength of the van der
Waals forces to be tuned, whilst maintaining all other parameters (protein and salt 
concentration, temperature, surface composition) constant. This 
concept has been successfully applied to describe the stability of coatings~\cite{karin_ralf2001} and 
has been recently extended to a comprehensive understanding  of gecko adhesion~\cite{gecko2005}.
An additional hydrophobization of the oxide surfaces by a self-assembled 
monolayer of silane molecules allows for a variation of the short-range 
forces, while maintaining the long-range forces essentially constant. The reason 
for this is that the strength of the van der Waals forces between an 
adsorbate and a layer are proportional to the volume of the layer~\cite{israel}.

Analysis of the kinetics facilitates characterization of protein adsorption
without the inherent assumptions often associated with the interpretation of
isotherms~\cite{norde1986}. Optical methods allow for high sampling 
frequency (in contrast to e.g.\ scattering techniques),
high resolution of the adsorbed amount, and {\it in-situ} monitoring without 
alteration of the protein structure or hydrodynamic conditions. 
Protein adsorption was followed {\it in-situ} with an imaging ellipsometer
(Nanofilm EP$^3$, Germany) operating via the nulling ellipsometry principle
at a wavelength of 532~nm~\cite{azzam1977}.
Samples were mounted in a teflon fluid cell which enabled precise temperature
and flow control at two angles of incidence, $65^\circ$ and $70 ^\circ$.
Modeling of the ellipsometric data assumes a homogeneous layer approximation, with de~Feijter's
method~\cite{defeijter1978} used to determine the adsorbed mass 
as both thickness and refractive index of such thin transparent films
cannot be unambiguously determined by single wavelength ellipsometry~\cite{azzam1977}.  
Assuming that the refractive index of
a protein in solution is a linear function of its concentration, the
absolute amount $\Gamma$ of adsorbed protein
can be determined by $\Gamma = d_f \frac{\left(n_f-n_a\right)}{dn/dc}$,
where $d_f$ and $n_f$ are the thickness and refractive index of the adsorbed film
respectively, $n_a$ is the refractive index of the ambient, and
$dn/dc$ is the refractive index increment of the molecules, which
was fixed at 0.183 cm$^3$/g for our measurements~\cite{defeijter1978,ball1998}.
The standard deviation is approximately $0.1$ mg/m$^2$.
Both wafer types (2~nm and 192~nm SiO$_2$) were hydrophilic and
were alternatively covered by a monolayer of silanes
(octadecyl-trichlorosilane), rendering them hydrophobic~\cite{wasserman1989,brzoska1994}.
This resulted in four different types of composite substrates with
pairwise identical short- or long-range forces, respectively~\cite{gecko2005,karin_ralf2001}.
The layer structure of all surfaces was characterized accurately by
{\it ex-situ} ellipsometry, the rms roughness of all surfaces was below $0.2$ nm
as determined by scanning probe microscopy of an area of $1$~$\mu \textrm{m}^2$.
$\alpha$-amylase from human saliva (Fluka no.~10092) was dissolved
in a 10~mM phosphate buffer solution at pH $7.0$, stored at 4 $^\circ$C, and used for up to 4 days 
after preparation. Upon the fluid cell reaching thermal equilibrium,
protein was injected under constant flow conditions (Rheodyne Manual Sample Injector),
which were maintained throughout a given experiment.

\begin{figure}[htbp]
  \centering
  \epsfig{file=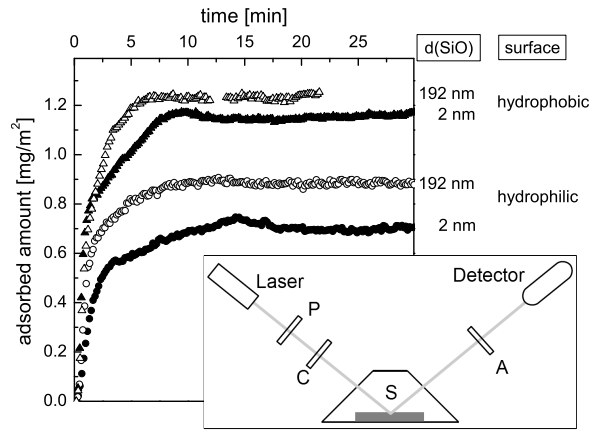,width=0.9\columnwidth,keepaspectratio}
  \caption{Adsorption of $\alpha$-amylase on the four substrates under
    investigation. 
    Inset: Ellipsometric setup in the PCSA configuration (P: Polarizer, C: Compensator, S: Sample, A: Analyser).
  }
  \label{fig:kinetics}
\end{figure}
Figure~\ref{fig:kinetics} shows the adsorbed amount of $\alpha$-amylase on the four substrates.
As expected from literature~\cite{malmsten2003},
we find a higher adsorbed amount on the hydrophobic surfaces.
The adsorption on the thick silicon oxide samples (open symbols) exhibits a
commonly observed kinetics with a continuously decreasing adsorption
rate up to a limiting value of the surface excess.
The situation is completely different on the native wafer series
(closed symbols). The initial adsorption rate is consistent with
that of the thermally grown oxide samples. However, one observes a
linear growth (constant growth rate) regime with a defined beginning
and end. This occurs on both the hydrophobic and hydrophilic surfaces.
The fact that one observes different adsorption kinetic curves on
chemically identical surfaces indicates that protein adsorption kinetics are
significantly influenced by long-range forces.
Additional experiments with various silicon oxide layer thicknesses
revealed that a three-step kinetics can be observed for SiO$_2$ layer
thicknesses below 20~nm.

\section*{The colloidal approach: \\ Model and Simulation results}

In order to explore the microscopic origin of the three-step kinetics 
we perform MC simulations using a colloidal representation of the protein 
molecules as spherical particles. 
The substrate and the solvent are likewise treated as continuous media.
Particle-particle as well as particle-surface interactions are described
in the framework of the DLVO-theory~\cite{lyophobic,lyophobic2},
considering steric repulsion, van der Waals and electrical double layer interactions. Hamaker's results~\cite{hamaker} are used to calculate the van der Waals 
forces.
Approximate expressions for the electrostatic interactions can be
obtained using the linear superposition approximation
(LSA)~\cite{lsa_idea} together
with Sader's equation~\cite{sader} for the effective far field potential.
Almost the complete set of model parameters, e.g.\ protein net charge and Debye
length, are experimentally accessible. Therefore we have chosen 
parameter values that are consistent with experimental
findings rather than model parameters optimizing the agreement
between experimental and theoretical results of the adsorption kinetics.  
\begin{figure}[htbp]
  \centering
  \epsfig{file=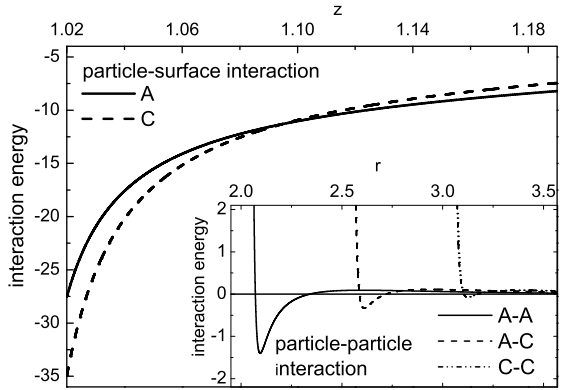,width=0.9\columnwidth,keepaspectratio}
  \caption{Conformation dependent particle-surface and particle-particle (inset) potentials.
    $z$ denotes the distance between the center of a particle and the substrate surface;
    $r$ is the center-to-center distance of two interacting particles. 
    Energies are given in units of $k_B T$, $z$ and $r$ in units of the radius of the spheres 
    in conformation~$A$.
  }
  \label{fig:potentials}
\end{figure}
In Fig.\ \ref{fig:potentials}, the solid curves represent the resulting 
particle-surface and (in the inset) particle-particle potentials.

The single particle MC scheme applied here describes the particle dynamics 
as a stochastic motion in real and configuration space, similar to the 
Brownian dynamics method. 
For this reason we expect to obtain qualitatively the same behavior
for the adsorption kinetics. 
However, the advantage of MC is that internal degrees of freedom of the particles can easily be considered,
enabling the complex adsorption behavior of proteins to be modeled.

\begin{figure}[htbp]
  \centering
  \epsfig{file=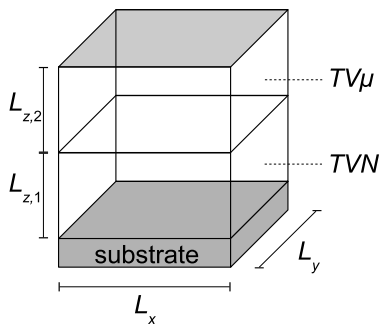,trim=0 145 0 0,clip,width=0.6\columnwidth,keepaspectratio}
  \caption{Sketch of the simulation box.
  }
  \label{fig:box}
\end{figure}

In order to investigate the time evolution of the surface coverage
theoretically it is necessary to choose a simulation volume that is
in accordance with the experimental situation. We have therefore
divided the simulation volume into two parts (see Fig.~\ref{fig:box})~\cite{lenhoff}.
In the upper box a grandcanonical ensemble ($TV\mu$) is applied. In
contrast, no particle exchange with an external reservoir is
considered in the lower box adjacent to the substrate.
Particles can diffuse from the upper box into the lower one and vice
versa. In the lower box they are influenced by the attractive
substrate such that there is a net particle flux from the upper box
into the lower one, until a stationary state is reached.
This setup is motivated by the experimental situation where the concentration
of proteins in bulk solution remains approximately constant during the
adsorption experiment. We use periodic boundary conditions in the $x$-
and $y$-direction and reflecting boundaries at the top and the bottom
of the simulation box.
The total height of the simulation box is chosen such that a further
increase of $L_{z,1}$ or $L_{z,2}$ does not change the simulation results qualitatively .

Using this basic model we observe the irreversible adsorption of a
particle monolayer consistent with the experimental findings.
As expected~\cite{lenhoff}, we obtain a conventional 
shape of the adsorption kinetics 
characterised by a gradual reduction of the adsorption rate.  
The robustness of this result indicates that the model has to be extended
in order to reproduce the observed three-step kinetics. 
In the experiments the long-range forces originating from the substrate
turn out to change the adsorption kinetics qualitatively.
A straightforward physical interpretation of this observation is that 
the long-range interactions influence the orientation of the proteins
in solution, and consequently their initial conformation in the irreversible 
adsorption process. This physical picture implies, that 
depending on the nature of the long-range forces, different initial
conformations of the adsorbed proteins are possible.

\section*{Modeling conformational changes}

As there is experimental and theoretical evidence 
that proteins in the adsorbed state
can undergo conformational changes \cite{zhdanov2001,castells2002,garcia2} 
it is crucial for the adsorption kinetics whether 
the conformation of adsorbed proteins is stable.
In order to maintain the computational performance,
we extended our model by introducing an internal degree of freedom,
which models the different protein conformations.
Specifically, particles may adopt three different states:
in bulk solution they take on the native (compact) conformation~$A$,
if adsorbed to the substrate they either adopt a marginally altered (native-like)
conformation~$B$ or a substantially altered (denatured) conformation~$C$.
Conformational changes $B \to C$ are modeled to be reversible 
in order to account for the process of partial refolding upon denaturation
($A \to B \to C \to B$)~\cite{zhdanov2001,castells2002}.
Trial probabilities are introduced for the thermally activated transitions $B \leftrightarrow C$.
As the effective radius of conformation~$B$
is assumed to be of the order of the native conformation~$A$, we choose $B \equiv A$ for simplicity.
The denaturation upon adsorption is associated with a larger
contact area between the protein molecule and the substrate (spreading), 
thereby enlarging the binding energy of the proteins, cf.\ Fig.\ \ref{fig:potentials}.
In the framework of the colloidal approach a conformational change
of a protein molecule upon adsorption is represented by a volume
conserving deformation of the particle. As a result the effective
size of the particle (``particle interaction radius'') with respect
to its interaction with the substrate is increased while it is
reduced with respect to the particle-particle interactions. In order
to maintain the computational efficiency it is favorable to keep
the spherical geometry of the particles. Consequently,
the denaturation is implemented by adapting the particle interaction radii
separately for the different contributions to the potentials (see
the inset of Fig.\ \ref{fig:potentials}). Note that particles in
conformation~$C$ now cover a larger surface area of the substrate.
Such a modeling of conformational changes is in the spirit of the
Equivalent Sphere Approach (ESA) for the interaction of non-spherical colloidal particles~\cite{esa1}.
The transition from compact to extended protein conformations takes place only
close to the surface of the substrate.
Therefore transitions $A \to C$ are restricted to particles that are in
close proximity to the substrate, i.e.\ for  $z \leq 1.1$,  where
conformation~$C$ is energetically favored (see Fig.\ \ref{fig:potentials}).
In contrast, conformational changes $C \to A$ 
are allowed in the entire simulation volume.
This ensures that unbound particles are in the native conformation~$A$.
Note that within the model presented a conformational change is indistinguishable from
a reorientation of a protein molecule at the surface (top-on $\leftrightarrow$ side-on).

\begin{figure}[htbp]
  \centering
  \epsfig{file=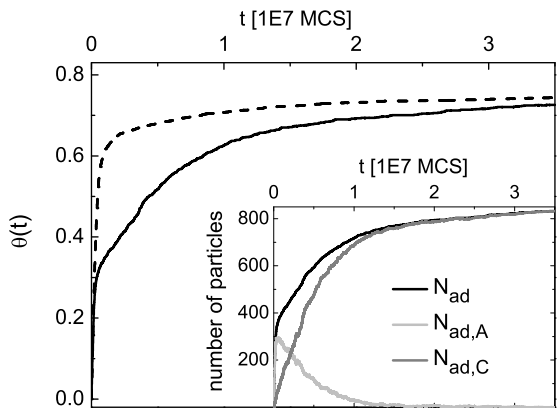,width=0.9\columnwidth, keepaspectratio}
  \caption{Adsorption kinetics for particles with (solid line) and
    without (dashed line) internal degree of freedom. Time~$t$ is
    measured in MC sweeps. Inset: Number of adsorbed particles depending on conformation.
  }
  \label{fig:kinetics_theory}
\end{figure}
Figure~\ref{fig:kinetics_theory} shows the time evolution of the
surface coverage~$\theta(t) = \frac{\pi}{L_{x} L_{y}} N_{ad}(t)$,
which is proportional to the number of adsorbed particles per surface area 
(for definition of $L_{x,y}$ cf.\ Fig.\ \ref{fig:box}) and corresponds to the experimentally
observed quantity adsorbed amount~$\Gamma$.
Compared to the reference curve (without conformational changes),
the adsorption kinetics is characterized by an intermediate region
with a moderate adsorption rate.  
\begin{figure}[tbp]
  \centering
  \epsfig{file=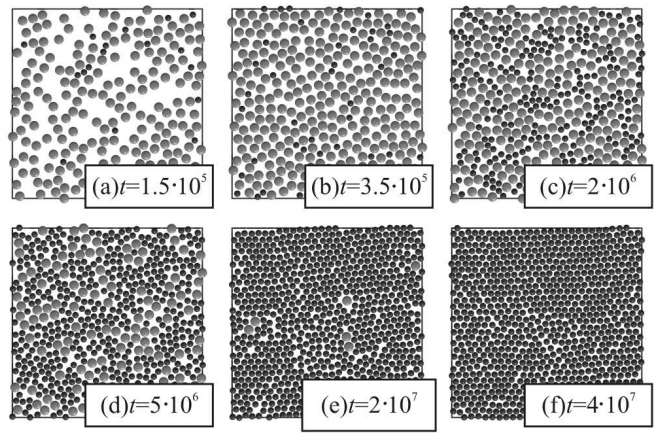,width=\columnwidth, keepaspectratio}
  \caption{Snapshots of the adsorption layer. The dark (bright)
    particles represent model proteins in conformation $A$ ($C$). 
  }
  \protect\label{fig:conf_mAB}
\end{figure}
In qualitative agreement with the experimental results
three regimes of the adsorption kinetics can be distinguished:
During the first part of the kinetics the
number~$N_{ad,C}$ of adsorbed particles in conformation~$C$ grows
almost as fast as the total number~$N_{ad}$ of adsorbed particles
(see inset of Fig.\ \ref{fig:kinetics_theory}), because at low
surface coverages the particles transform immediately after
adsorption to the energetically favored conformation~$C$
(see Fig.\ \ref{fig:conf_mAB}(a),(b)). 
With increasing surface coverage, the particle-particle interactions
become more relevant and induce the growth of $A$-domains at the
surface (see Fig.\ \ref{fig:conf_mAB}(c),(d)).
For these high surface coverages the optimization of particle-particle
interactions due to the formation of $A$-domains overcompensates the unfavorable
surface-particle interaction of conformation~$A$. The third step
of the adsorption kinetics can be viewed as the ordering transition of a 2d
monodisperse system (see Fig.\  \ref{fig:conf_mAB}(e),(f)).
This process leads to a rather slow saturation of the adsorption kinetics
compared to the experimental observations. However, for the real system
a rearrangement to a closest packed structure is not expected anyway,
since at high surface coverages, the entanglement of proteins
plays an important role, which is not considered in our model.

Thus, according to the model presented, the occurrence of the
discontinuity and the second linear regime in the adsorption
kinetics can be ascribed to a collective transition in the internal
degree of freedom of the particles, namely from a conformation that
is stable on the single-particle level~($C$) to a conformation that
optimizes adsorbed amount at the surface~($A$). Discrepancies of the
simulated adsorption kinetics for large times are inherent to the
colloidal model.

The collective transition is observed for a wide range of realistic
model parameters. Whether or not the effect of surface-induced
conformational changes leads to a second linear regime in the
adsorption kinetics depends on how fast the collective transition
takes place. The swiftness of the transition is given by the
decrease of $N_{ad,C}$ and is influenced by several simulation
parameters, e.g.\ the particle density, the time scale of the
internal degree of freedom, the ratio of the effective particle
radii and the relative strength of the conformation-dependent
particle-particle and particle-surface interactions.

\section*{Conclusions}

In conclusion, evidence for the dependence of the adsorbed
amount of protein on short-range interactions (hydrophilic/hydrophobic) as
well as on the long-range van der Waals forces  
is demonstrated.
This is contrary to the common belief that protein adsorption is predominantly determined
by the short-range contribution of the surface potential.

Surprisingly, modifications of the long-range interactions
result in a three-step adsorption kinetics.
The simulation results indicate
that this kind of kinetics originates from conformational changes of
the proteins at the surface which are induced
by density fluctuations. 

In order to confirm the physical picture developed here, it would be
ideal to perform experiments that can directly probe the conformations
of the adsorbed proteins. 
Unfortunately, it is not possible to obtain this kind of information {\it in-situ} 
with state-of-the-art experimental methods. 
In addition to ellipsometry, methods are available that
enable the adsorbed protein layer to be characterized in more detail,
but these methods suffer from either a lack of time resolution for {\it in-situ}
experiments (e.g.\ neutron scattering~\cite{fragneto1995nrs}) or from a constraint 
concerning the variation of substrate composition (e.g.\ surface plasmon resonance spectroscopy~\cite{green2000spr}).
Establishing quantitative agreement between experimental and simulation results
requires to refine the model approach. 
However, the experimental time scales and the large number of molecules 
restrain the complexity of a protein model, which can still be used to simulate
biofilm adsorption with state-of-the-art computers. 
So far it is not even possible to simulate 
the denaturation upon adsorption of a single protein molecule in full atomistic detail,
thereby gaining access to its surface native state~\cite{ganazzoli}.
A feasible route to improve systematically the level of sophistication of the colloidal model
is to include structural information provided by coarse-grained molecular dynamics simulations~\cite{tozzini2005}. 

Nevertheless, our observations show that long-range van der Waals forces, 
emananting from the substrate bulk material,
may alter the properties of adsorbed protein films.
Hence, for a comprehensive study of protein adsorption it 
is of great importance
to take this type of interaction into account. \\ \\

\begin{acknowledgments}

The authors thank M.\ Hannig and K.\ Huber for inspiring
discussions and gratefully acknowledge the support of the Alexander von Humboldt
and the German Science Foundation under grant number SA 864/2-2,
as well as Wacker Chemitronics AG and Nanofilm GmbH for technical support.

\end{acknowledgments}

\end{document}